\documentclass[fleqn,usenatbib]{mnras}

\usepackage{newtxtext,newtxmath}

\usepackage[T1]{fontenc}

\usepackage{graphicx}	
\usepackage{amsmath}	

\newcommand{\Msun}{\ensuremath{\text{M}_\odot}}

\newcommand{\sSFR}{\ensuremath{sSFR}}
\newcommand{\Mstar}{\ensuremath{M_*}}
\newcommand{\lsSFR}{\ensuremath{{\log(\frac{\sSFR}{\text{yr}^{-1}})}}}

\newcommand{\lMstar}{\ensuremath{{\log(\frac{\Mstar}{\Msun})}}}

\newcommand{\dd}{\text{d}}
\newcommand{\deriv}[2]{\frac{\dd{#1}}{\dd{#2}}}
\newcommand{\pp}{\ensuremath{\deriv{ \,P(\,\lsSFR\,|\,\Mstar) }{ \,\lsSFR }}}

\newcommand{\pcond}{\ensuremath{\,P(\,\lsSFR\,|\,\Mstar)}}

\title[Do galaxies die?]{Do galaxies die? Different views from simulations and observations in the Local Universe}

\author[P. Corcho-Caballero et al.]{
Pablo Corcho-Caballero,$^{1,2}$\thanks{E-mail: pablo.corcho@uam.es}
Yago Ascasibar,$^{1}$
Cecilia Scannapieco$^{2}$
\\
$^{1}$Departamento de Física Teórica, Universidad Autónoma de Madrid (UAM), Campus de Cantoblanco, Madrid 28049, Spain\\
$^{2}$Australian Astronomical Optics, Macquarie University, 105 Delhi Rd, North Ryde, NSW 2113, Australia\\
$^{3}$Universidad de Buenos Aires, Facultad de Ciencias Exactas y Naturales, Departamento de F\'{\i}sica. Buenos Aires, Argentina\\
}

\date{ Accepted 2021 June 29. Received 2021 June 29; in original form 2021 May 10}

\pubyear{2021}

\begin{document}
\label{firstpage}
\pagerange{\pageref{firstpage}--\pageref{lastpage}}
\maketitle

\begin{abstract}
For years, the extragalactic community has divided galaxies in two distinct populations.
One of them, featuring blue colours, is actively forming stars, while the other is made up of ``red-and-dead'' objects with negligible star formation.
Yet, are these galaxies really dead?
Here we would like to highlight that, as previously reported by several independent groups, state-of-the-art cosmological numerical simulations (EAGLE, IllustrisTNG, MAGNETICUM and SIMBA) predict the existence of a large number of quenched galaxies that have not formed any star over the last few Gyr.
In contrast, observational measurements of large galaxy samples in the nearby Universe (GAMA, SDSS) suggest that even the most passive systems still form stars at some residual level close to $sSFR\sim10^{-12}~\text{yr}^{-1}$.
Unfortunately, extremely low star formation poses a challenge for both approaches.
We conclude that, at present, the fraction of truly dead galaxies is still an important open question that must be addressed in order to understand galaxy formation and evolution.
\end{abstract}




\section{Introduction}

Galaxies are often classified into two distinct populations according to their specific star formation rate, i.e. the instantaneous star formation rate $SFR=\dot M_*$ per unit stellar mass $M_*$, $sSFR=SFR/M_*$.
``Passive'' systems, also referred to as ``red-and-dead'' galaxies, are supposed to form stars at a much lower (or even negligible) rate compared to ``star-formig'' galaxies (SFG) \citep[e.g.][]{2010Peng, 2015Peng, 2021Katsianis}.
However, this picture has been criticized by several recent studies \citep{2017Feldmann, 2017Eales, 2018Eales, 2020Corcho}, that challenge the existence of such a bimodal galaxy population.

Most SFG are well known to arrange along a relatively narrow region in the $M_*-sSFR$ plane dubbed the ``Main Sequence'' (MS) of star-forming galaxies\ \citep[e.g.][]{2007Noeske, 2011Elbaz, 2015Renzini, 2015Sparre, 2019Popesso}, where the sSFR roughly corresponds to the inverse of the age of the Universe; $sSFR\sim10^{-10}~\text{yr}^{-1}$ at the present time.
As recently argued in\ \citet{2020Corcho}, galaxies are not distributed along a one-dimensional relation of the form $sSFR(M_*)$.
Instead, the conditional probability distribution \pp\ is found to be well described by a simple analytical ansatz with only one maximum, tracing what has traditionally been identified as the MS, and relatively shallow power law tails that account for the starburst and passive populations.
In any case, the full conditional probability distribution of galaxies across the $M_*-sSFR$ plane provides much more information than its moments $\langle sSFR(M_*) \rangle$ and $\sigma_{sSFR}(M_*)$, or the global sSFR function $\deriv{N}{sSFR}(sSFR)$ integrated over all stellar masses.

Here we are concerned with the precise location of passive galaxies on the $M_*-sSFR$ plane, confronting the insight provided by cosmological numerical simulations and observations of galaxies in the local Universe.
The traditional view of a bimodal galaxy population \citep{2006Dekel, 2008vandenBosch, 2010Peng, 2015Peng} posits the existence of one or more ``quenching'' processes that transform them into red-and-dead systems by truncating star formation on timescales of the order of $\sim0.1-1$ Gyr \citep[e.g.][]{2013Wetzel, 2019Phillipps,2019Wright}.
Recently, the dichotomy between living and dead galaxies has become a major concern among the extragalactic community, leading to multiple definitions of the ``passive'' regime: a uniform threshold $sSFR<10^{-11} ~\text{yr}^{-1}$, 1 $\sigma$ deviation from the MS $sSFR(M_*)<\text{MS}(M_*)-\sigma(MS)$, or colour-based cuts \citep{2019Davies, 2019Donnari}.
Several groups have focused on the question of whether passive galaxies
are effectively detached from the star-forming population or they form a continuous distribution with gradually decreasing levels of star-formation \citep{2017Feldmann, 2017Eales, 2020Corcho}.

Unfortunately, accurately measuring any level of star formation is always challenging, and intrinsic differences between observational tracers have been reported in the literature\ \citep{2007Salim, 2012Kennicutt, 2016Davies, 2020Katsianis}. 
This fact becomes critical at the low-SFR regime, both for low-mass galaxies with faint fluxes as well as massive systems with young stellar fractions of the order of one percent or below\ \citep{2020Salvador-Rusinol, 2020deLorenzo-Caceres}. 
Besides, selection effects at different wavelengths may affect the statistical distribution of galaxies if not properly accounted for\ \citep{2018Eales, 2019Davies}.
On the other hand, the results of numerical simulations are still sensitive to resolution and subgrid physics, and strong convergence is still not achieved\ \citep{2015Schaye, 2020Zhao}.

Previous studies have recently hinted statistically significant differences between the distribution of observed and simulated sSFRs in passive systems\ \citep{2020Zhao, 2021Katsianis}.
Simulated galaxies may naturally be divided into SFG ($SFR>0$) and ``quenched'' ($SFR=0$) systems, while observed ``passive'' galaxies form stars at a rate $sSFR\sim10^{-12}~\text{yr}^{-1}$, not obviously detached from the SFG.

It is at present unclear whether empirical measurements are affected by systematic and/or statistical biases nor whether simulated passive galaxies are realistic, but addressing this question is of the utmost importance in order to understand the physical mechanisms that regulate the SFR.
Here we make a thorough statistical comparison between state-of-the-art cosmological simulations and large observational surveys in the local universe, paying special attention to the distinction between passive and quenched galaxies.
Section~\ref{sec:data} describes the observational and numerical data sets upon which the present work is based.

\section{Data}
\label{sec:data}

\subsection{Observational data}
\label{sec:obs_data}

We used two galaxy samples, taken from Sloan Digital Sky Survey (SDSS) DR7 \citep{2009Abazajian} and Galaxy and Mass Assembly (GAMA) DR3 \citep{2018Baldry}, restricted to the local Universe ($0.001<z<0.1$).
As both surveys are flux limited, we applied the following selection criteria, based on Petrosian photometry, to ensure completeness:
\begin{center}
$
 \textnormal{SDSS}~\left\{ \begin{array}{c} m_u<19.0 \\m_g<19.0 \\m_r<17.7                                            \end{array}\right.;~
 \textnormal{GAMA}~\left\{ \begin{array}{c} m_u<23.0 \\m_g<23.0 \\m_r<19.8    \end{array}\right.
$   
\end{center}

SDSS SFRs and stellar masses were taken from the MPA-JHU spectroscopic catalog\ \citep{2003Kauffmann, 2004Brinchmann, 2004Tremonti}. 
With these constraints, the present sample of SDSS comprises the 2.5, 16, 50, 84 and 97.5 percentiles of stellar mass and SFR of 162,279 galaxies.
Stellar masses were computed from a Bayesian analysis of the $ugriz$ photometry based on a set of star formation templates including exponential star formation histories with random bursts, applying minor corrections for nebular emission using fibre spectra.
For those galaxies with enough S/N, not classified as AGN hosts, SFRs were derived within the fibre aperture by fitting emission line measurements -- H$\alpha$, H$\beta$, [OIII]5007, [NII]6584, [OII]3727, [SII]6716 -- with galaxy evolution models combining stellar population synthesis and emission line modelling under a Bayesian framework.
SFRs of AGN hosts or galaxies with no emission lines are estimated using the 4000\AA~Balmer break, previously calibrated with the high-S/N sample. 
Aperture corrections are applied using the galaxy photometry following \ \citet{2007Salim}.
For more details regarding the computation of SFRs see \citet{2004Brinchmann}.

\begin{figure*}
    \centering
    \includegraphics[width=0.485\linewidth]{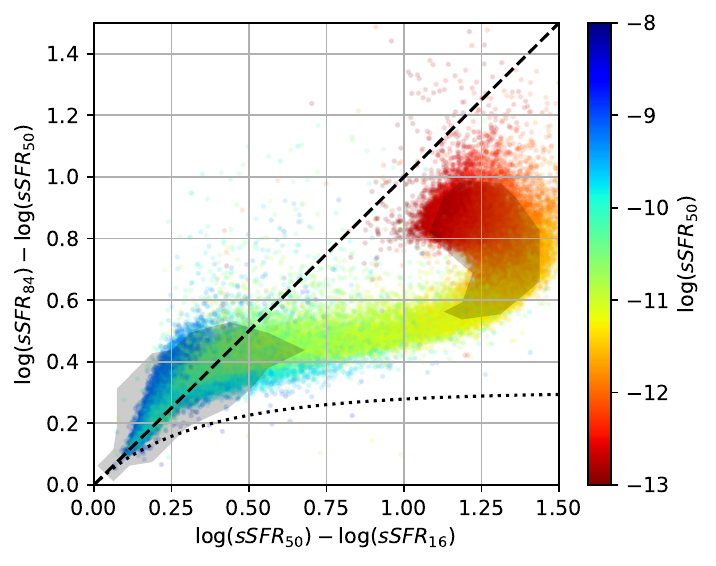}
    \includegraphics[width=0.485\linewidth]{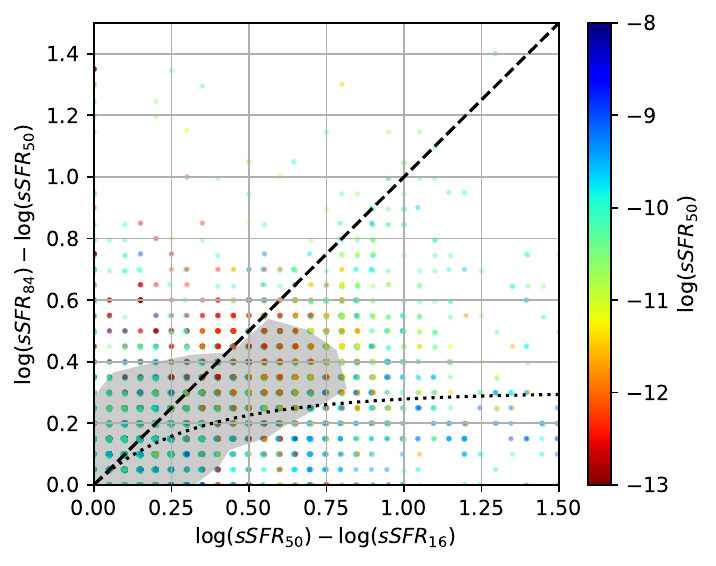}
    \caption{Difference between different percentiles of individual sSFR PDFs in SDSS (\textit{left}) and GAMA (\textit{right}) data, colored by the 50 percentile.
    Black dashed lines illustrate the log-symmetric case, and dotted lines denote the linear-symmetric scenario.
    Grey shaded areas enclose 95 percent of each sample.
    }
    \label{fig_extra:observational_errors}
\end{figure*}
    
The percentile distributions of GAMA SFRs and stellar masses were computed by fitting panchromatic spectral energy distributions based on 21 photometric bands in the wavelength range $0.1$-$500~\mu$m\ \citep{2016Wright} using the energy balance code \texttt{MAGPHYS}\ \citep{2008dacunha}, assuming exponential star formation histories across a broad range of characteristic time scales.
For a detailed description see\ \citet{2016Driver, 2018Driver}.

We have accounted for the uncertainties using the reported percentiles in $M_*$ and $sSFR$, and we applied the 1/V$_\textnormal{max}$ correction method in order to estimate the number density of galaxies in the ${M_*-sSFR}$ plane.
We emphasize that accurately measuring low levels of star formation is extremely challenging. 
For this reason, instead of using the median values reported for each galaxy, we computed their individual probability density distribution (PDF): 

\begin{equation}
\begin{split}
    \rho_i(\,\lMstar,\,\lsSFR\,) = \deriv{P_i(\lsSFR)}{\lsSFR}&\\ \times\deriv{P_i(\lMstar)}{\lMstar},
\end{split}
\end{equation}

where $P_i$ is reconstructed by linearly interpolating the published 2.5, 16, 50, 84 and 97.5 percentiles in order to approximate the PDF of both $M_*$ and $sSFR$, neglecting any possible covariance\ \citep{2020Corcho}.
Therefore, the total probability distribution is nothing but the sum of all individual elements in the sample:

\begin{equation}
\deriv{^2 P}{\,\lMstar\,\dd\lsSFR}
= \frac{ \sum_i^N \rho_i / V_{\text{max}, i} }{ \sum_i^N 1 / V_{\text{max}, i} }.
\end{equation}

Let us stress that the individual probability distributions are far from being symmetric, especially at low sSFRs.
We plot on Figure~\ref{fig_extra:observational_errors} the difference between the 50 and 16 percentiles versus the difference between the 84 and 50 percentiles for each galaxy in our SDSS and GAMA samples, colored by their median $sSFR_{50}$ values.
For SDSS data, MS galaxies tend to display moderate, log-symmetric errors, of the order of $\sim 0.1-0.5$~dex, whereas passive systems cluster around $sSFR_{84}-sSFR_{50} \sim 0.8$ and $sSFR_{84}-sSFR_{50} \sim 1.25$, and galaxies in the green valley scatter along a continuous distribution between both extremes.
Note that only the MS galaxies show a log-symmetric probability distribution.
Such correlation between $sSFR$ and asymmetry is totally absent in GAMA data, with typical uncertainties below $~\sim0.6$~dex.

The precise shape of the probability distribution is critical in order to estimate the fraction of observed galaxies compatible with $SFR=0$.
According to the published measurements, the probability of the null value is always below the $2.5$ percentile.
Unfortunately, this result is extremely sensitive to the adopted priors, and we strongly advocate that a rigorous statistical analysis should be carried out specifically to address this important question.

\begin{table}
    \centering
    \begin{tabular}{|c|c|c|c|}
        \hline
        Label $^a$ & Run $^b$  & $L_{\text{box}}^c$ & $m_g^d$ \\
        \hline
         EAGLE100 & Ref-L0100N1504 & 100 & $1.81\times10^{6}$ 
         \\
         EAGLE50 & Ref-L0050N0752 & 50 & $1.81\times10^{6}$ 
         \\
         EAGLE25 & Ref-L0025N0752 & 25 & $2.26\times10^{5}$ 
         \\
         TNG300 & TNG300-1 & 302 & $1.1\times10^7$
         \\
         TNG100 & TNG100-1 & 111 & $1.4\times10^6$
         \\
         MAG500 & Box2 & 520 & $1.4\times10^8$
         \\
         SIMBA150 & m100n1024 & 148 & $1.82\times10^7$
         \\
         SIMBA75 & m50n1024 & 74 & $2.28\times10^6$
         \\
         SIMBA35 & m25n1024 & 37 & $2.85\times10^5$
         \\
         \hline
    \end{tabular}
    \caption{Simulation suites. 
    $^a$ Label adopted in this work.
    $^b$ Original run name given by the collaboration.
    $^c$ Box length in comoving Mpc.
    $^d$ Gas element resolution in \Msun.
    }
    \label{tab:simus}
\end{table}

\subsection{Cosmological simulations}
\label{sec:simu_data}

We selected several state-of-the-art hydrodynamical cosmological simulations based on a variety of numerical codes.
By employing cubic periodic volumes that extend from tenths to hundreds of Mpc per side, it is possible to track the evolution of galaxies from the very early stages of the Universe until the present time.
Baryonic processes acting at sub-resolution scales (e.g. star formation, feedback from stars and active galactic nuclei)
are computed through sub-grid schemes based on different physical models, whose free parameters are fixed by reproducing observables like the galaxy stellar mass function, the galaxy size-mass relation or the black hole-halo mass relation\ \citep{2020Vogelsberger}.
All the simulations used in the present work take into account radiative cooling, heating from ultraviolet background radiation, star formation, feedback from SF, black hole growth and feedback from AGN.
The following paragraphs specify the particular runs that have been selected from each project.
Adopted labels, original run names, box sizes and mass resolutions are summarised in Table~\ref{tab:simus} for the sake of comparison.

\textbf{EAGLE.} This set of simulations  were run using a modified version of the Smoothed Particle Hydrodynamics (SPH) code \texttt{GADGET}3\ \citep{2015Schaye, 2015Crain}.
We use three runs, named by the collaboration as Ref-L0100N1504, Ref-L0050N0752 and Ref-L0025N0752 (hereafter, EAGLE100, EAGLE50 and EAGLE25) with box lengths of 100, 50 and 25 comoving Mpc, respectively.
The gas density threshold for a gas particle to be classified as ``star forming'' is $n_H= 0.1~\text{cm}^{-3}$, and its instantaneous SFR is computed following a pressure-dependent Kennicutt-Schmidt law \citep{2015Schaye}.

\textbf{IllustrisTNG.} It constitutes the next generation of the Illustris project \citep{2014Vogelsberger}, that uses the moving-mesh \texttt{AREPO} code \citep{2018Pillepich, 2018Springel, 2018Nelson, 2018Marinacci, 2018Naiman}. 
We have used runs TNG100-1 and TNG300-1 (hereafter TNG100 and TNG300), that share the same default model parameters with side lengths of $\sim$ 100 and $\sim$ 300 Mpc, respectively \citep[][]{2019Nelson}.
Star  formation  and  pressurization  of  the  multi-phase  interstellar  medium are  treated  according to \citep{2003Springel}.
Gas particles above a density threshold $n_H= 0.1~\text{cm}^{-3}$ form stars stochastically following the empirical Kennicutt-Schmidt relation.

\textbf{MAGNETICUM.} \citep{2014Hirschmann, 2017Ragagnin}
This set of simulations were run with the SPH code \texttt{GADGET}3.
We use the high-resolution Box2 run, that extends $\sim$500 Mpc per side (hereafter, MAG500). 
Star  formation  and  its associated feedback processes also follow\citep{2003Springel}.

\textbf{SIMBA.} 
This suite inherits the main framework of the MUFASA simulations \citep{2016Dave}, based on the \texttt{GIZMO} code \citep{2019Dave}.
We use three different runs, dubbed by the collaboration as ``full physics'' models: m25n1024, m50n1024, m100n1024, with side lengths $\sim$35, $\sim$70 and $\sim$150 Mpc, respectively (hereafter, SIMBA35, SIMBA75 and SIMBA150).
To form stars these simulations employ a molecular based star formation prescription following \citep{2009Krumholz} and require a minimum density of $n_H=0.13~\text{cm}^{-3}$.

For all simulation suites, stellar masses are computed as the sum of all star particles bound to each galaxy, and total SFRs are the sum of the individual instantaneous SFR of all gas particles/cells in each galaxy.
We restricted this study to galaxies with stellar masses in the range $8.5<\log(M/M_\odot)<11.5$.
Since both $M_*$ and $sSFR$ are directly computed from the simulations, no statistical uncertainties are involved.

\section{Results}
\label{sec:results}

\subsection{Probability distribution}

\begin{figure*}
    \centering
    \includegraphics[width=.485\linewidth]{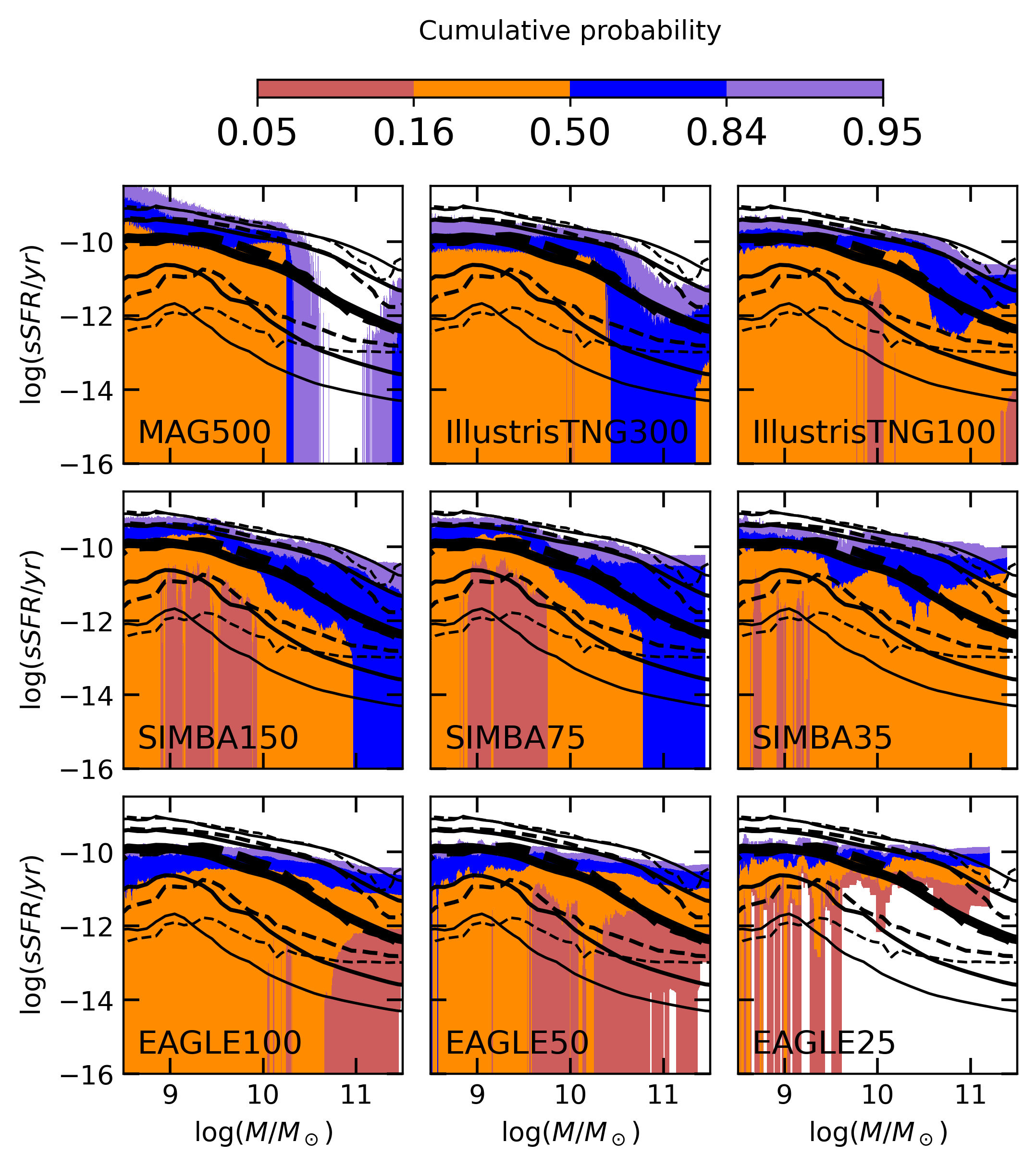}
    \includegraphics[width=.485\linewidth]{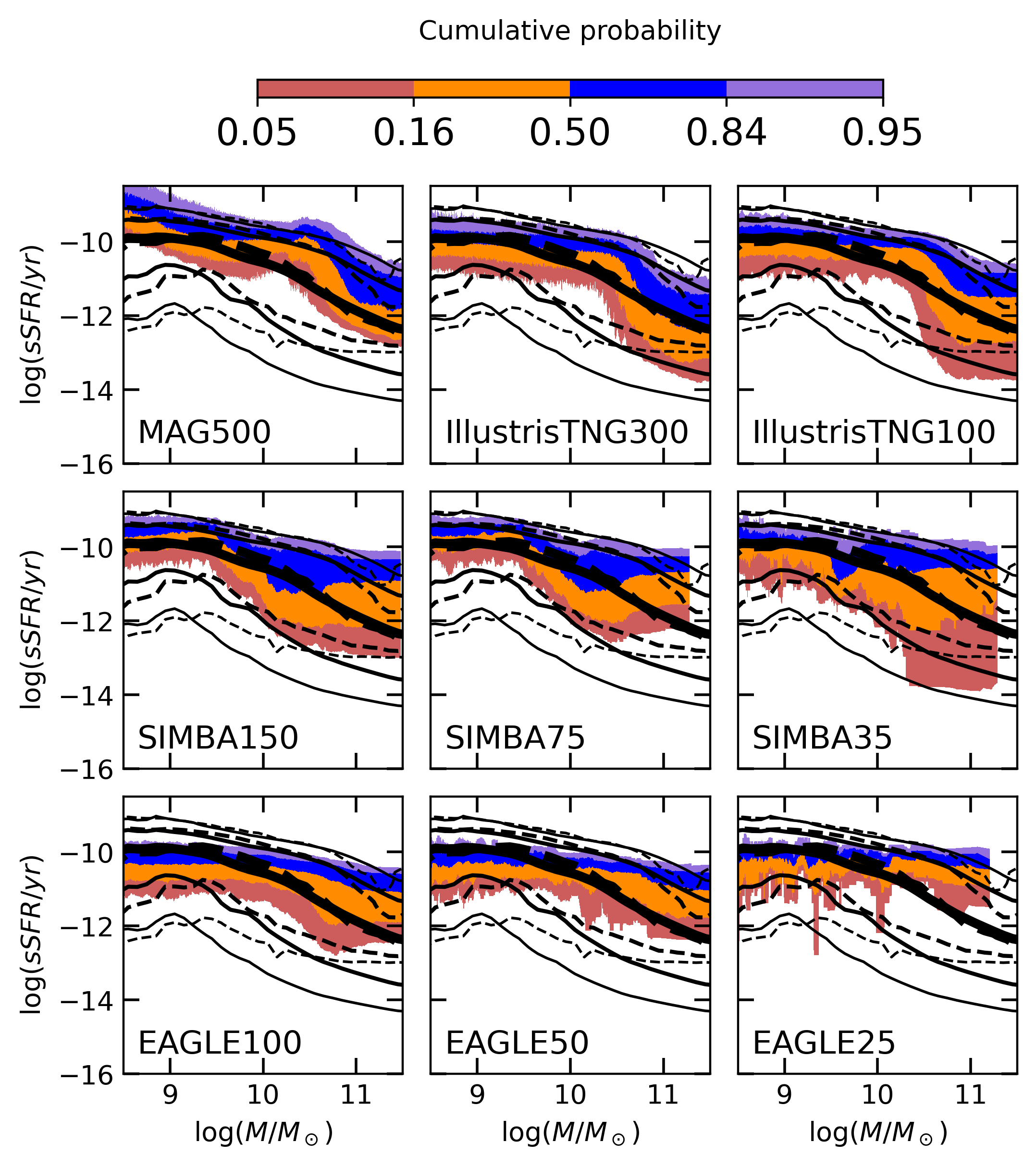}
    \caption{Cumulative probability distribution $\pcond$.
    For each simulation run there are two panels: considering all galaxies (\textit{left}) and excluding galaxies with $SFR=0$ (\textit{right}). 
    Colored contours denote the cumulative probability for simulated data. Black lines denote SDSS (solid) and GAMA (dashed) 50 (thickest), 84 and 16 (mid-thickness), 5 and 95 (thinnest) percentiles.}
    \label{fig:cumulative_prob}
\end{figure*}

\begin{figure*}
    \centering
    \includegraphics[width=.8\linewidth]{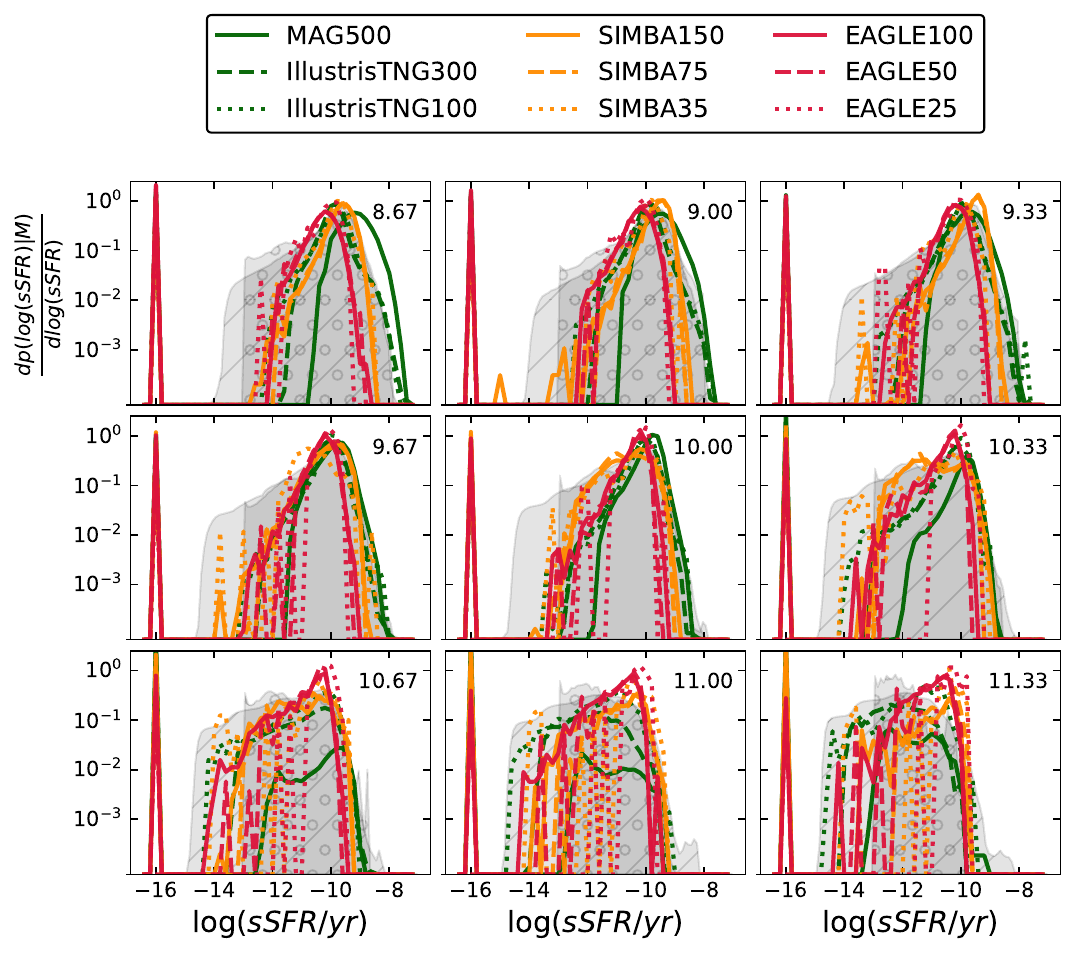}
    \caption{Conditional probability of $sSFR$ in 9 mass bins. The number on the upper-right corner of each panel denotes the $\log(M/M_\odot)$ bin mid value, with 0.33 dex width. Shaded regions correspond to SDSS and GAMA distributions (line-hatched and circled-hatched, respectively). Each colored line corresponds to different simulations runs (see legend).}
    \label{fig:density_prob}
\end{figure*}

Let us start by showing that the sSFR of the ``passive'' population is very different in numerical simulations and in observations of galaxies in the local Universe.
In order to illustrate this discrepancy, we have computed the cumulative conditional probability $\pcond$ for the set of publicly available cosmological simulations and observational samples described in Section~\ref{sec:data}.
Figure~\ref{fig:cumulative_prob} displays 9 panels corresponding to different simulation runs from the EAGLE, IllustrisTNG, MAGNETICUM and SIMBA projects (colored contours), compared with two observational samples from SDSS and GAMA (solid and dashed lines, respectively). 
Quenched galaxies (QG) with $SFR=0$ are excluded on the right panels, and they have been assigned a value $sSFR=10^{-16}~\text{yr}^{-1}$ in the left panels for representation purposes (a similar value is adopted in all other Figures).

We first notice that, when excluding quenched galaxies, the simulated MS is roughly in agreement with the observed one, albeit significantly narrower.
Our results support the lack of simulated galaxies with low (yet non-zero) sSFR with respect to the observed distribution reported by \citet{2020Zhao, 2021Katsianis}.
The difference is much larger for SDSS than for GAMA data, which might be explained by the larger observational errors.

In most cases, including QG into the sample does not have a major effect on the high $sSFR$ tail of the distribution.
However, the lowest percentiles are deeply affected.
Depending on the prescriptions adopted by each code, the median may remain stable at low masses, but substantial changes are observed in the high-mass regime in all cases except EAGLE25.
In some particular instances, abrupt changes are observed above a certain critical mass, in stark contrast with the observed distribution.
The discrepancy in the 5 and 16 percentiles often becomes dramatic, and none of the codes under study is able to reproduce the observed distribution over the whole mass range.

There is also relevant information that can be extracted from the conditional probability density distribution \pp.
On Figure~\ref{fig:density_prob} we have divided each sample in 9 mass bins to illustrate the discrepancy.
Observational results appear as shaded regions with diagonal lines (SDSS) and circles (GAMA).
Taking uncertainties into account, GAMA data hint a possible bimodal distribution where passive galaxies would display $sSFR\sim10^{-12}~\text{yr}^{-1}$, in contrast to SDSS.
Both distributions are statistically compatible with a unimodal distribution with power-law tails that extend in both directions \citep{2020Corcho}, but it is clear that in any case there would be a significant overlap between the passive and star-formation populations, if they were distinct at all.
Changing the number of mass bins does not significantly affect the main conclusions extracted above.

On the other hand, all simulations (represented by coloured lines) consistently feature a remarkable fraction of quenched galaxies with $SFR=0$, clearly detached from the star-forming population.
Although the normalisation may be severely affected in a few cases, there is overall good agreement with observations near the star-forming peak \citep{2017Katsianis, 2020Zhao, 2021Katsianis}.
Simulated galaxies are naturally divided into two discrete populations: it is interesting to note that star-forming systems, excluding QG, are well described by a unimodal continuous probability distribution from starburst to passive galaxies akin to the one proposed in \citep{2020Corcho}.
In contrast, most QG completely ceased star formation several Gyr ago (see section \ref{sec:simulated_sfr}).


\subsection{Low-redshift sample}

If quenched galaxies really exist, it is important to address whether they would be detected by large galaxy surveys like SDSS or GAMA, according to their limitations.
At the low-mass end, especially regarding the quenched satellite population, it is likely that many of these systems would not form part of the observed sample, either due to their faint fluxes or owing to their close distance to the central companion.

To accomplish a more thorough comparison we used the photometric catalogues available for the IllustrisTNG and EAGLE suites to compute the maximum distance at which each galaxy would be detected by both surveys based on their synthetic photometry.
The sample used here only includes EAGLE and IllustrisTNG suites, that provide catalogues for $ugriz$ broadband magnitudes.
For each band, the maximum distance for a galaxy to fall within the detection limits is given by
\begin{equation}
d_{max}/\text{Mpc} = 10^{ \frac{m_{i, lim}-m_{i}}{5} - 5 }
\end{equation}
where $m_i$ denotes the absolute magnitude of the galaxy in the $i\in\{u,g,r\}$ band, $m_{i, lim}$ corresponds to the limit imposed to select the observational sample, and the minimum aver all bands is adopted to assign a maximum distance $d_{max}$ to each galaxy.

\begin{figure}
\centering
    \includegraphics[width=\linewidth]{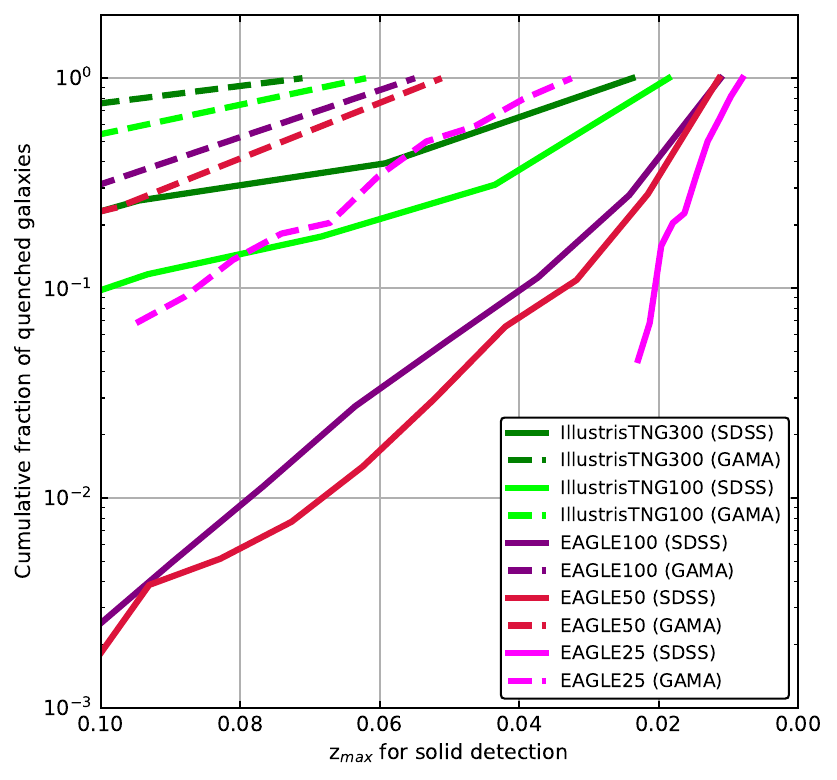}
    \caption{Survey detection. Cumulative fraction of quenched (passive) galaxies as function of the maximum redshift for detection in SDSS and GAMA surveys.}
    \label{fig_extra:quenched_detection}
\end{figure}

Figure~\ref{fig_extra:quenched_detection} shows the fraction of QG that would be detected in SDSS (solid lines) and GAMA (dashed) up to redshift $z_{max}$.
Solid lines denote the results obtained with SDSS selection limits and dashed lines illustrate GAMA constrains.
Consequently, it is very likely that our current SDSS observational sample ($z<0.1$) is missing most QG systems.
On the other hand, GAMA constraints allow a detection of roughly the whole population of quenched galaxies at $z\leqslant0.05$. 

\begin{figure}
\centering
    \includegraphics[width=\linewidth]{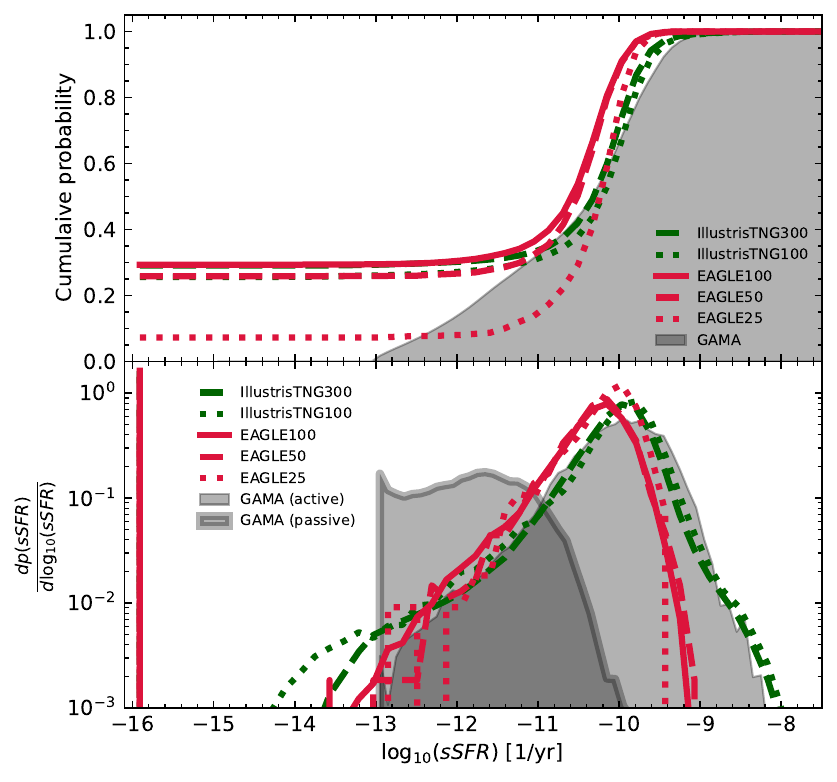}
    \caption{Probability distributions at $z\leqslant0.05$. (\textit{left}) Cumulative probability distribution $P(sSFR<sSFR_i)$. Grey shaded region denotes observational data from GAMA (restricted to $z\leqslant0.05$) and colored lines correspond to simulation runs (see legend). (\textit{right}) Probability density distribution of $sSFR$. Grey shades denote the GAMA star-forming (non-contoured) and passive (black contoured) distributions ($z\leqslant0.05$).}
    \label{fig:ssfr_distrib}
\end{figure}

Thus, we have compared the simulated probability distribution of the sSFR with a subsample of GAMA galaxies restricted to this redshift range.
The cumulative probability can be seen on the top panel of Figure~\ref{fig:ssfr_distrib}.
Above $sSFR\approx10^{-11}$ the agreement is quite reasonable for the majority of runs, meaning that the total fraction of passive galaxies is well predicted, as discussed by \citet{2020Donnari}.
However, simulations lack passive systems with low levels of star formation, that are replaced by QG.
As an alternative proof, we show on the bottom panel of this figure the probability density distribution assuming a bimodal population in GAMA.
We divided the GAMA sample into star-forming and passive systems by means of the traditional threshold $sSFR_{50}=10^{-11}$~yr$^{-1}$ referred to the measured median value.
We find that the star-forming distribution (grey shaded area), broadened by the observational uncertainties, matches well the whole simulated PDF of galaxies with $SFR>0$.
This can be interpreted in terms of a correspondence between the passive population of GAMA (contoured shaded region) and the concentrated peak of QG in the simulated distributions.

This is the root of the discrepancy; there is good agreement above $sSFR=10^{-11}$~yr$^{-1}$, but there is a significant mismatch on the star formation activity of the passive systems, as previously noticed by other studies \citep[e.g.][]{2020Zhao, 2021Katsianis}.
The present results provide strong evidence that this discrepancy is not due to selection effects in observational samples.

\subsection{Simulated sSFR}
\label{sec:simulated_sfr}

\begin{figure*}
    \centering
    \includegraphics[width=0.49\linewidth]{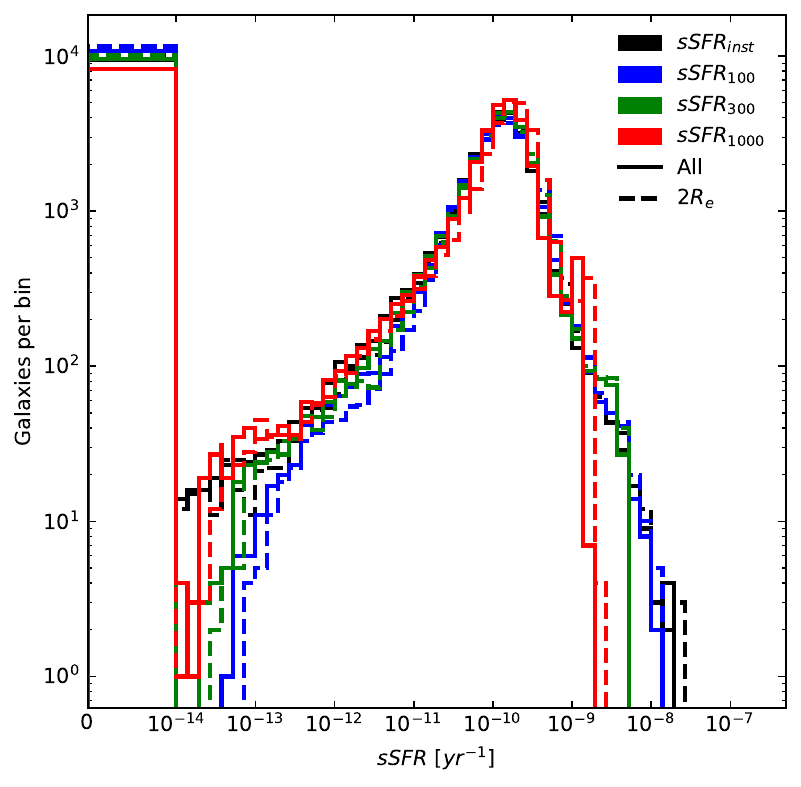}
    \includegraphics[width=0.49\linewidth]{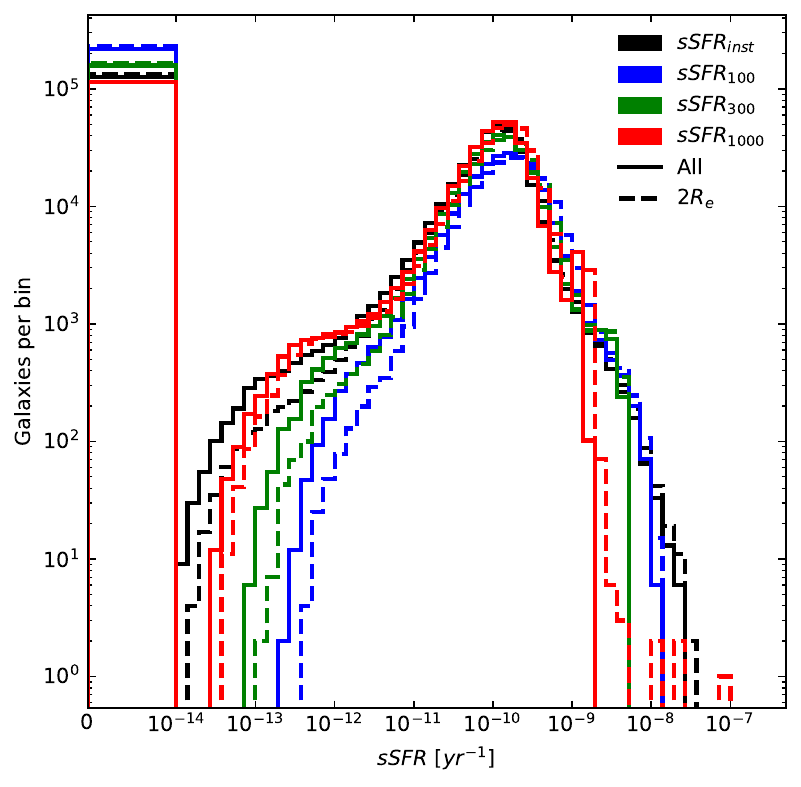}
    \caption{IllustrisTNG sSFR distributions for TNG100 (\textit{left}) and  TNG300 (\textit{right}). Colored lines denote different $sSFR$ timescales: instantaneous (black), 100 Myr (blue), 300 Myr (green) and 1000 Myr (red). Continuous lines represent values computed for the whole extent of each galaxy while dashed lines denote values restricted to two effective radii.}
    \label{fig_extra:illustrisTNG_ssfr_distrib}
\end{figure*}

\begin{figure*}
    \centering
    \includegraphics[width=0.49\linewidth]{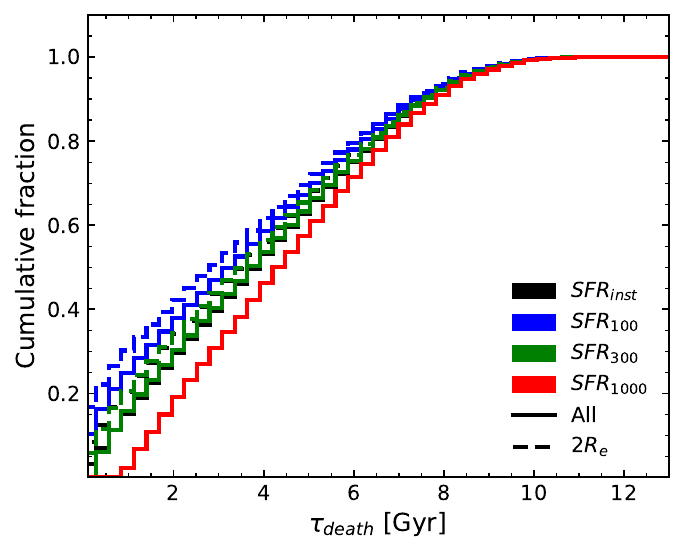}
    \includegraphics[width=0.49\linewidth]{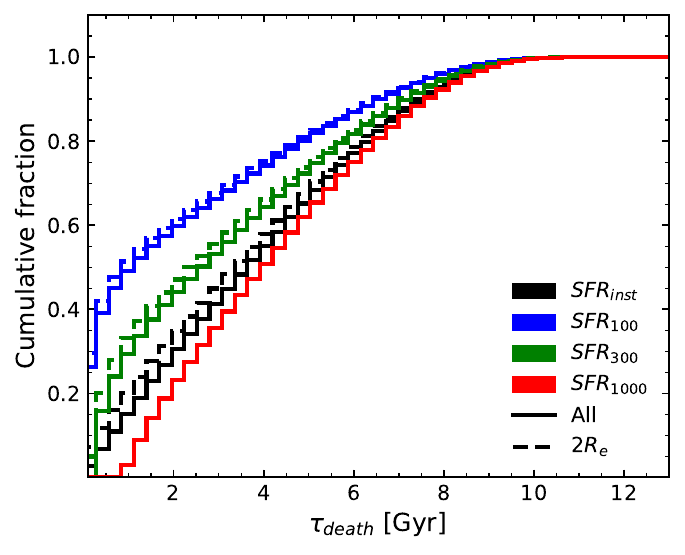}
    \caption{Age of the youngest stellar particle, $\tau_{\rm death}$, for TNG100 (\textit{left}) and  TNG300 (\textit{right}). Each line denotes cumulative fractions derived according the same scheme as shown in Fig~\ref{fig_extra:illustrisTNG_ssfr_distrib}.}
    \label{fig_extra:dead_times}
\end{figure*}

Other potential systematics are explored in Figure~\ref{fig_extra:illustrisTNG_ssfr_distrib}.
Observationally derived star formation rates are effectively averaged over some particular timescale that ranges from tenths to hundreds of Myr, depending on the proxy used to trace the fraction of young stellar populations, whereas our default simulated SFRs represent the instantaneous value.
Here we also consider the amount of stellar mass formed during the last 100 Myr ($SFR_{100}$), 300 Myr ($SFR_{300}$) and 1000 Myr ($SFR_{1000}$).
In particular, the $SFR_{100}$ timescale matches the multiwavelength analysis of GAMA data \citep{2016Davies}. 
Furthermore, observations might only cover the inner parts of the galaxies, and therefore we will also analyse the star formation rate restricted to two effective radii.

The overall shape of the distribution is not affected at all by the choice of indicator.
We find that aperture effects are negligible in practice for our present purposes.
At the high-SFR end, averaging over Gyr scales suppresses the most intense starbursts, whereas using the average star formation within a short time interval instead of the instantaneous SFR is limited by the mass of one individual stellar particle.
As a result, the number of quenched galaxies (i.e. $SFR=0$) is slightly larger, but the qualitative prediction remains the same.

The minimum values of the SFR that we find in each simulation of the IllustrisTNG suite \citep[see also appendix A of][for a detailed discussion]{2019Donnari} are $\log(SFR_{inst}) = \{-3.53,\ -2.81\}$, $\log(SFR_{100}) = \{-2.20,\ -1.81\}$, $\log(SFR_{300})=\{-2.64,\ -2.32\}$, and $\log(SFR_{1000})=\{-3.12,\ -2.58\}$ for TNG100 and TNG300, respectively.
Even with time-averaged SFRs, these simulations would be able to confidently resolve $sSFR < 10^{-11}$~yr$^{-1}$, at least for the most massive galaxies.
Using the instantaneous star formation rate, or its average over the last Gyr, QG are clearly distinct from passive galaxies with low sSFR, irrespective of the mass resolution of the simulation, although their predicted number might be sensitive to this parameter (as well as box sizes potentially sampling different environments).


In addition, we have also checked that the population of QG is not due to stochastic variations of the SFR.
As shown in Figure~\ref{fig_extra:dead_times}, they have been dead, in the sense of $SFR=0$, for a significant fraction of the age of the universe.
Half of them ``died'' about $3-4$~Gyr ago and did not form any stellar particle ever since.

\begin{figure}
    \centering
    \includegraphics[width=\linewidth]{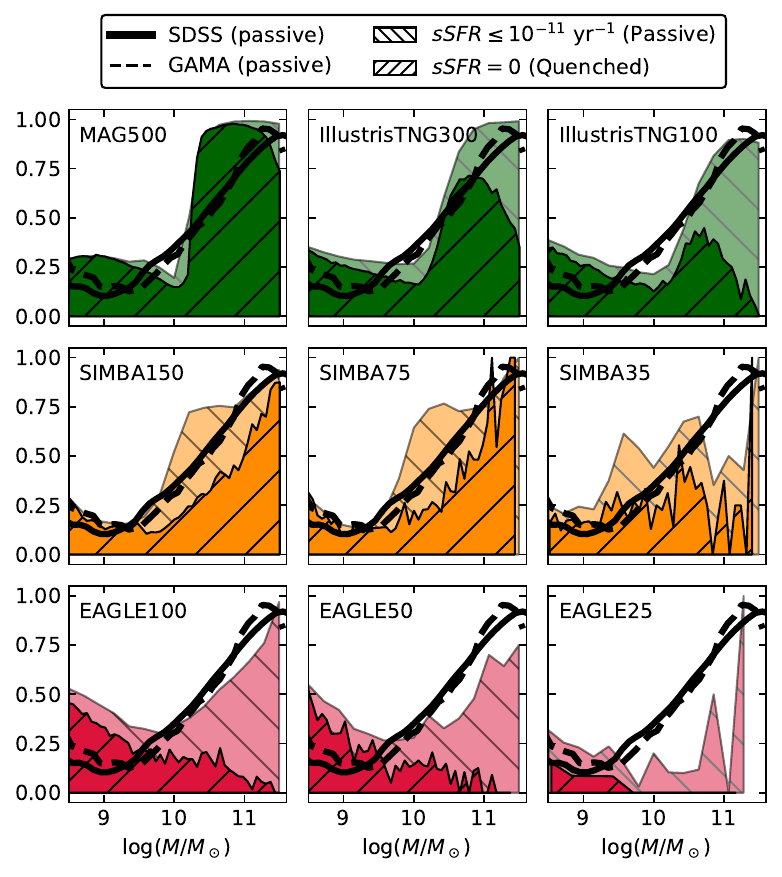}
    \caption{Quenched and passive fraction. Relative fraction of passive/quenched galaxies in simulations as function of total stellar mass. Black solid and dashed lines denote the SDSS and GAMA passive fraction ($sSFR\leq10^{-11}$) respectively.}
    \label{fig_extra:quenched_fraction}
\end{figure}

\subsection{Passive and quenched fractions}

On Figure~\ref{fig_extra:quenched_fraction} we have represented the relative fraction of passive ($sSFR<10^{-11}$ yr$^{-1}$) and QGs ($sSFR=0$) as a function of total stellar mass for all the simulations used in this work. 
We also show the passive fraction found in SDSS (solid black line) and GAMA (dashed black line) samples.
When considering passive galaxies, simulated and observed fractions  reasonably agree. 
However, this agreement is reached due to the presence of a large number of systems with $SFR=0$, included in the passive population, that may even dominate the total passive fraction.
Quenched galaxies roughly represent $\sim20-40$ percent of the dwarf population, with the fraction decreasing systematically with mass.
On the intermediate and high-mass regime, when reaching a critical mass $\sim10^{10}~M_\odot$, the fraction raises drastically in most cases, reaching values as high as $\sim50-90\%$.
Once again, we find that numerical resolution might play a role regarding the precise fraction of QGs at a given stellar mass.
In particular, EAGLE, IllustrisTNG and SIMBA suites always display a lower fraction of quenched galaxies at higher resolution.
We also find substantial differences depending on the prescriptions adopted by the different algorithms.

\section{Discussion and conclusions}
\label{sec:conclusions}

Our results are consistent with previous studies reporting that the observed and simulated distributions coincide above  $sSFR=10^{-11}$~yr$^{-1}$, and both approaches yield a similar fraction of systems below that threshold \citep{2020Donnari}.
Discrepancies in the distribution of passive galaxies with $SFR > 10^{-1.5}$~\Msun~yr$^{-1}$ have been highlighted by several authors.
\citet{2020Zhao} argue that a stronger quenching would be necessary for simulations to reproduce the debated \citep{2020Corcho} peak in the sSFR distribution at $\sim 10^{-12}$~yr$^{-1}$.
Our results are more consistent with the opposite view \citep{2021Katsianis} that quenching in numerical simulations should be \emph{weaker}, so that galaxies maintain a low star formation activity rather than becoming completely extinguished.

An alternative scenario would be that the origin of this ``dead galaxies problem'' could be an overestimation of the measured SFR below a certain threshold.
The ionising radiation used to trace recent star formation episodes could also come from other sources aside of young stellar populations \citep{2020Sanchez}, and
galaxies with reported $sSFR\sim10^{-12}$~yr$^{-1}$ might be simply dominated by such emission being in fact dead.
This remains an open question, and deeper analysis from both approaches will be necessary to solve it.

From the results presented here, we advocate for including the full conditional probability distribution in the $M_*-sSFR$ plane as an essential test to compare simulations and observations.
This is particularly interesting in order to test feedback processes and baryonic physics in general, that are possibly responsible for regulating not only the overall fraction of galaxies with low $sSFR$ but also their precise levels of star formation activity.
On the observational side, a more detailed analysis of the passive population must be performed.
This issue may not affect some other key aspects or fundamental relations, but its relevance is crucial in order to establish a robust theory of galaxy evolution.

We can conclude that current observational data are inconsistent with cosmological simulations within the reported error bars.
Unfortunately, systematic uncertainties on both sides make it difficult to clearly pinpoint the origin of the problem and provide a solution.
On the one hand, simulations successfully reproduce many of the observed galaxy properties, but feedback processes are far from being fully understood.
The precise fraction of dead galaxies as a function of stellar mass is still sensitive to numerical resolution and sub-grid physics (Figure~\ref{fig_extra:quenched_fraction} below), but the existence of a significant population of dead galaxies with $SFR=0$ seems to be consistently predicted.

On the other hand, current galaxy surveys provide a wealth of valuable information about galaxy demographics, but limitations in data analysis (e.g. the choice of priors when using Bayesian statistics, model degeneracies, selection effects) may significantly bias the results.

The stellar masses and SFRs used in the present work were inferred by means of analytical star formation histories, covering a wide range of different parameters.
However, this approach rests on a number of assumptions (e.g. dust attenuation law, galactic chemical evolution) that may severely affect the accuracy of the derived properties \citep{2020Lower}.
In particular, it is implicitly assumed in this methodology that galaxies do \emph{not} die, and it is mathematically impossible to measure a null star formation rate at any time.

Nevertheless, there have been previous studies that have explicitly investigated the fraction of galaxies in different environments that show statistical evidence for a sudden change in their star formation history \citep{2016Boselli, 2018Ciesla, 2020Aufort}.
According to \citep{2020Aufort}, over 90 per cent of the galaxy population can indeed be well described by a smooth, positive star formation history.
About one third of the galaxies that show evidence for sudden variations of the SFR with respect to a delayed exponential model correspond to recent star formation bursts, whereas the other two thirds display significantly decreased (but non-zero) star formation activity after some quenching event that took place in the recent past.
Therefore, the lack of a significant population of dead galaxies in observational measurements seems to be robust with respect to the assumed star formation history.

Let us note that the question of whether galaxies die or not is closely related to the location, or even existence, of a ``passive'' peak in the conditional probability distribution of the sSFR as a function of mass.
At present, it is very hard to conclude whether:
\begin{enumerate}
\item There is a significant population of galaxies that did not form any star during the past few Gyr.
\item There is a bimodal distribution of sSFR, with passive galaxies featuring typical values of the order of $\sim 10^{-12}$~yr$^{-1}$.
\item There is a unimodal conditional probability distribution as a function of stellar mass, with a power law tail towards low sSFR \citep{2020Corcho}.
\end{enumerate}
These scenarios are qualitatively different, and discriminating between them has crucial implications on the underlying physics, but additional research is clearly needed in order to address this question.

\section*{Acknowledgements}

P.C. and Y.A. acknowledge financial support from the Spanish Government project ESTALLIDOS: PID2019-107408GB-C42 (Ministerio de Ciencia e Innovaci\'on, Spain). 
C.S. acknowledges financial support from Agencia Nacional de 
Investigaciones Cient\'{\i}ficas
y Tecnol\'ogicas (Argentina), through PICT-206-0667.

\section*{Data Availability}
 
All data that support the findings of this study are publicly available.
The MPA-JHU spectroscopic catalog is available at \url{https://www.sdss.org/dr12/spectro/galaxy_mpajhu/}.
The catalog of galaxy properties derived from the GAMA panchromatic survey is available at \url{http://www.gama-survey.org/dr3/schema/table.php?id=82}.
The following links correspond to the web pages where simulations data can be accessed.
EAGLE \url{http://icc.dur.ac.uk/Eagle/database.php}. 
IllustrisTNG \url{https://www.tng-project.org/data/}.
MAGNETICUM \url{http://www.magneticum.org/data.html#FULL_CATALOUGES}
SIMBA \url{http://simba.roe.ac.uk/}.

The codes that support the plots within this
article are freely available for download at \url{https://github.com/PabloCorcho/do_galaxies_die}.



\bibliographystyle{mnras}
\bibliography{references} 







\bsp	
\label{lastpage}
\end{document}